\documentclass{moriond}

\usepackage{wrapfig}
\usepackage{lipsum}
\usepackage{float}
\usepackage{subcaption}
\def\Journal#1#2#3#4{{#1} {\bf #2}, #3 (#4)}

\def\be{\begin{equation}}
\def\ee{\end{equation}}
\def\bea{\begin{eqnarray}}
\def\eea{\end{eqnarray}}

\newcommand{\Photo}{\includegraphics[height=35mm]{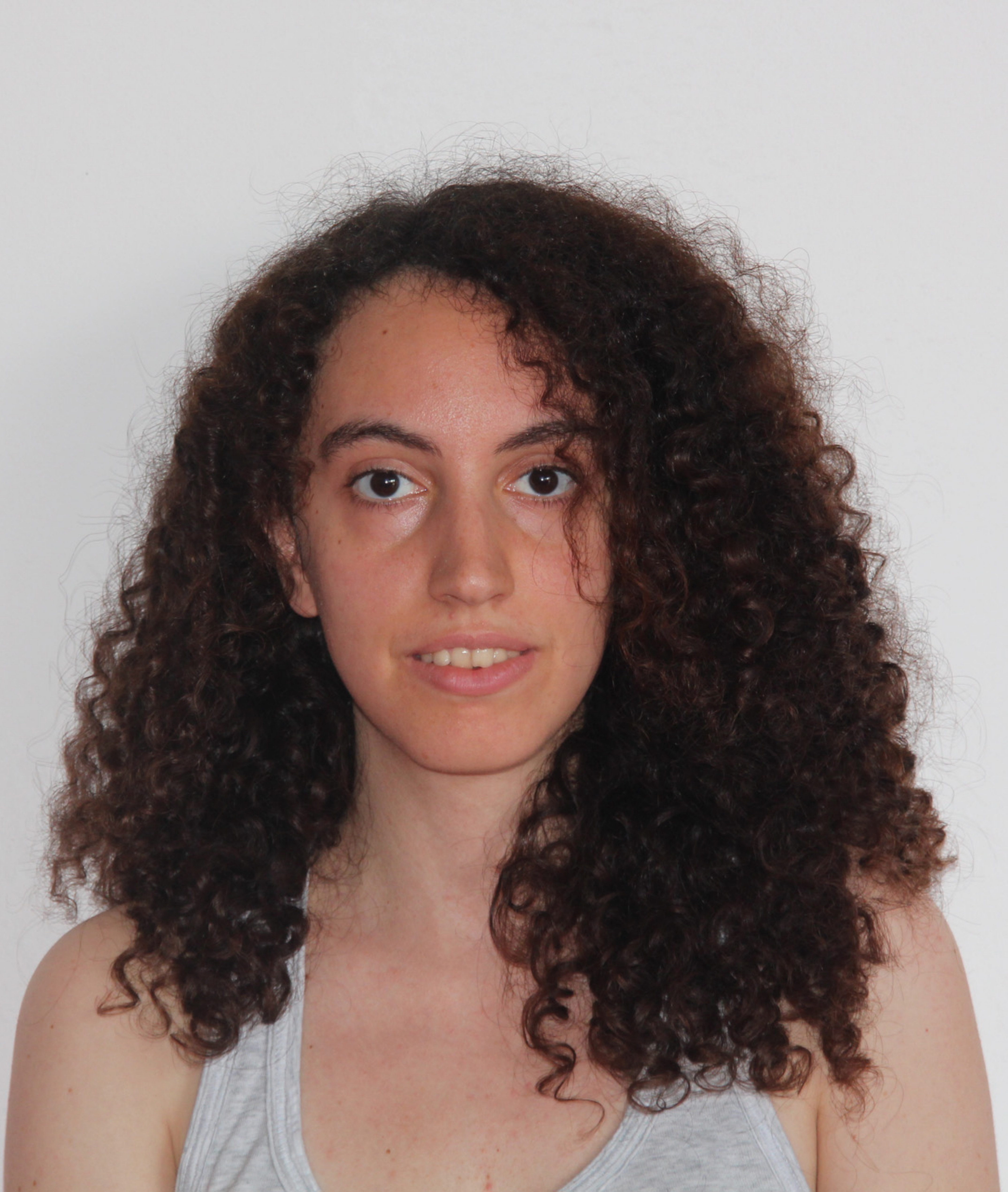}}

\begin{document}
\vspace*{4cm}
\title{DARK MATTER SEARCHES TOWARD THE GALACTIC CENTER HALO WITH H.E.S.S.}

\author{\underline{Lucia Rinchiuso} and Emmanuel Moulin for the H.E.S.S. Collaboration}

\address{IRFU, CEA, Universit\'e Paris-Saclay, F-91191 Gif-sur-Yvette, France}

\maketitle\abstracts{
The presence of dark matter in the Universe is nowadays widely supported by a large body of astronomical and cosmological observations.
The central region of the Milky Way is expected to harbor a large amount of dark matter. Very-high-energy ($>$100 GeV) $\gamma$-ray observations with the H.E.S.S. array of Imaging Atmospheric Cherenkov Telescopes are powerful probes to look for self-annihilations of dark matter particles toward the Galactic Centre. A new search for a dark matter signal has been carried out on the full H.E.S.S.-I dataset of 2004-2014 observations with a 2D-binned likelihood method using spectral and spatial properties of signal and background. Updated constraints are derived on the velocity-weighted annihilation cross section for the continuum and line dark matter signals. Higher statistics from the 10-year Galactic Center dataset of H.E.S.S.-I together with a novel analysis technique allow to significantly improve the previous limits. }

\section{Dark matter at the Galactic Center}

\subsection{Why dark matter?}\label{subsec:prod}
Many evidences support the presence of dark matter (DM), from cosmological observations to galaxies dynamics. The idea of DM was introduced in the `30s to explain the galaxy rotation curves that suggested a lack of mass with respect to the observed luminosity. Furthermore, the gravitational lensing proved the presence of matter which does not emit light but deviates the light that travels in its vicinity. More recently, thanks to measurements of the cosmic microwave background, the relic density of cold dark matter (CDM) has been estimated to be $\Omega_{\rm CDM}h^2=0.1186\pm0.0020$ \cite{bib:PDG}: about 25$\%$ of the Universe is made of DM. In the standard model of cosmology, dark matter is expected to be ``cold'', {\it i.e.} non-relativistic, to explain the scale of the formation of large scale structures. Among the most promising particle candidates for dark matter is non-baryonic weakly interacting massive particles (WIMPs) \cite{bib:wimp}, with a mass and couplings of the order of the electroweak scale.


\subsection{Gamma-ray flux from dark matter annihilations}
In dense astrophysical environments, DM particles are expected to pair annihilate. Among the final products very high energy (VHE, E$>$100 GeV) $\gamma$-rays can be found. These photons are not bent by magnetic fields at galactic scale and thus point back to the source. The flux of photons from DM self-annihilation can be computed as $$\frac{d\phi(\Delta\Omega,E)}{dE}=\frac{1}{4\pi}\frac{\langle\sigma v\rangle}{2m_{\rm DM}^2}\sum Br_{\rm i}\frac{dN_{\rm i}(E)}{dE}\times J(\Delta\Omega).$$
A particle physics term accounts for the DM mass $m_{\rm DM}$, the thermally-averaged velocity-weighted annihilation cross section $\langle\sigma v\rangle$, and the sum over all the spectra $\frac{dN_{\rm i}}{dE}$ in the different annihilation channels weighted by their branching ratio $Br_{\rm i}$. An astrophysics term, referred as to the J-factor, corresponds to the integral of the DM density squared over the line of sight (los) and integrated over the solid angle $\Delta\Omega$: $$J(\Delta\Omega)=\int_{\Delta\Omega}\int_{\rm los}\rho^2(r(s,\theta)dsd\Omega.$$ The coordinate $r$ is $r(s,\theta)=(r_\odot^2+s^2-2r_\odot s\cos\theta)^{1/2}$ and the solid angle writes $d\Omega=\sin\theta\cos\theta d\phi$, where $s$ is the coordinate along the line of sight, $\theta$ the angle between the direction of observation and the GC plane and $r_\odot=8.5$ kpc is the distance of the GC to the Sun. Two kinds of $\gamma$-ray spectra are expected from DM pair-annihilation: a continuum and a line spectrum. The hadronization and/or decay of the annihilation products such as W$^{\pm}$/Z bosons, quarks and leptons, produce a continuum spectrum of $\gamma$-rays with an energy cut-off at the DM mass. This secondary production of $\gamma$-rays constitute the dominant channel. Leptonic channels have very hard spectra while   hadronic ones have the maximum 
\begin{wrapfigure}{l}{0.55\textwidth} 
  \begin{center}
    \includegraphics[width=0.5\textwidth]{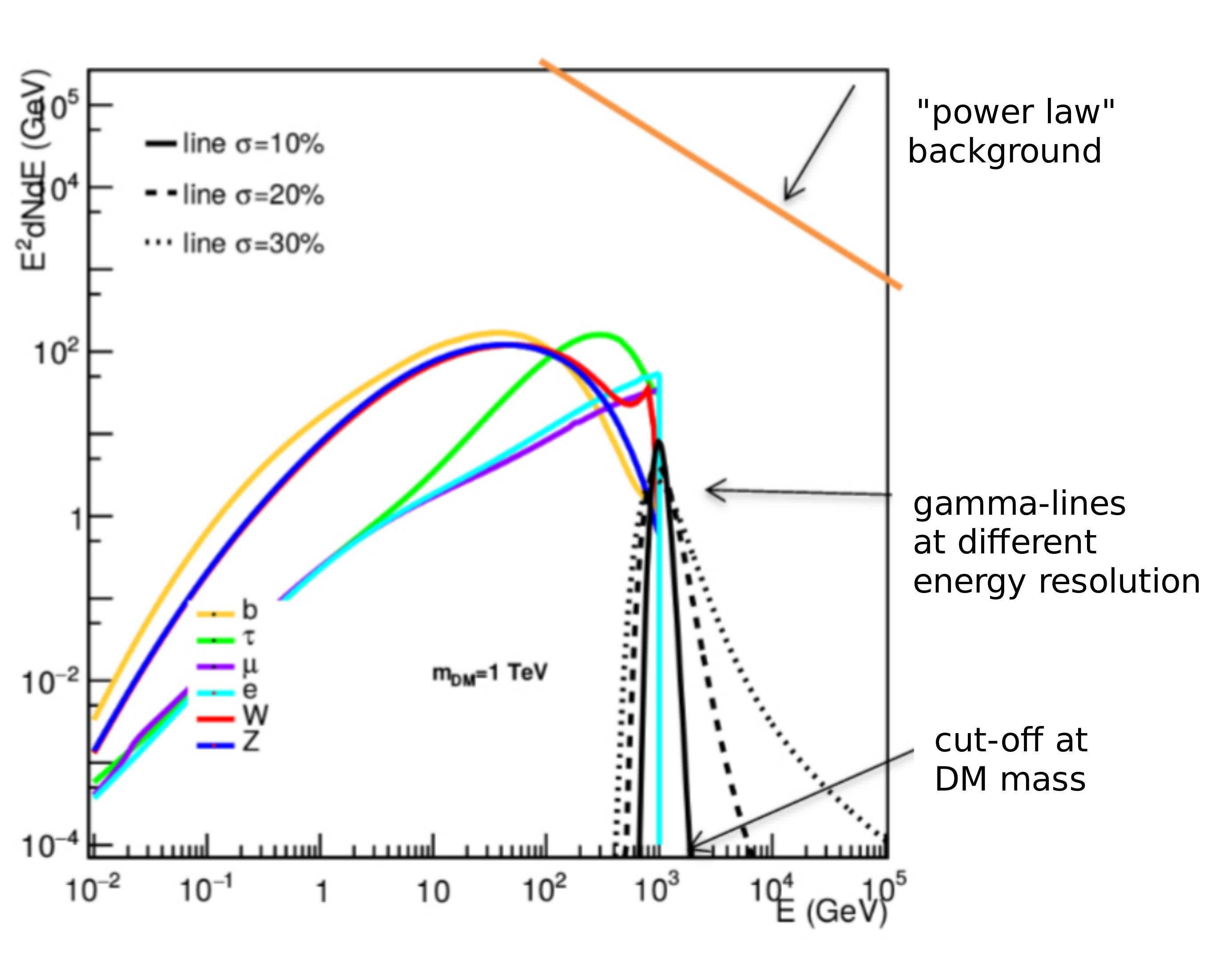}
    \caption{DM annihilation spectra for the continuum signal and $\gamma$-line for $m_{DM}=1$ TeV. The line (in black) is spread with different energy resolutions (10, 20 and 30$\%$). The ''power law'' background is represented in orange.}
    \label{fig:spectra}
  \end{center}
\end{wrapfigure} 
at lower energies. In presence of charged particles ({\it e.g.} in the $W^+W^-$ channel) in the annihilation process the $\gamma$-ray spectrum can also be modified by electroweak (EW) effects such as final state radiation and virtual internal bremsstrahlung contributions \cite{bib:EW}. The $\tau^+\tau^-$ channel shows both the features of the pure hadronic and leptonic channels. The line spectrum corresponds to the prompt annihilation into two photons. This process is suppressed compared to the continuum emission because it cannot take place at the tree-level. However, its peculiar spectrum is the clearest signature of DM annihilation as it can be efficiently discriminated against astrophysical background emissions. A line spectrum is obtained also when the photon is produces in pair or with a Z or Higgs boson. The latter process is not accounted for in this study. Using the measurements of the thermal relic density $\Omega_{\rm CDM}$ with the Planck satellite, the natural scale for $\langle\sigma v\rangle$ is of about $3\times10^{-26}$ cm$^3$s$^{-1}$ for the continuum signal, and $\sim10^{-29}$ cm$^3$s$^{-1}$ for the line signal, respectively.
$\gamma$-ray spectra of different channels are plotted in Fig. \ref{fig:spectra}. 
The EW corrections in the $W^+W^-$ (red line) channel induce an enhancement near the DM mass. The effect of the energy resolution on the line signal is also shown (black lines).
The orange line is intended to show the power law-like behavior of the astrophysical background.
Therefore the spectral information can be exploited to improve the discrimination between signal and background. 

The DM distribution in the Milky Way is not well constrained at distance less than about 1 kpc from the GC. Several density profiles have been proposed. Fig. \ref{fig:DMprofile} shows two kinds of profiles : 
\begin{figure*}[htbp]
    \centering
    \begin{subfigure}[t]{0.45\textwidth}
        \centering
        \includegraphics[height=2.1in]{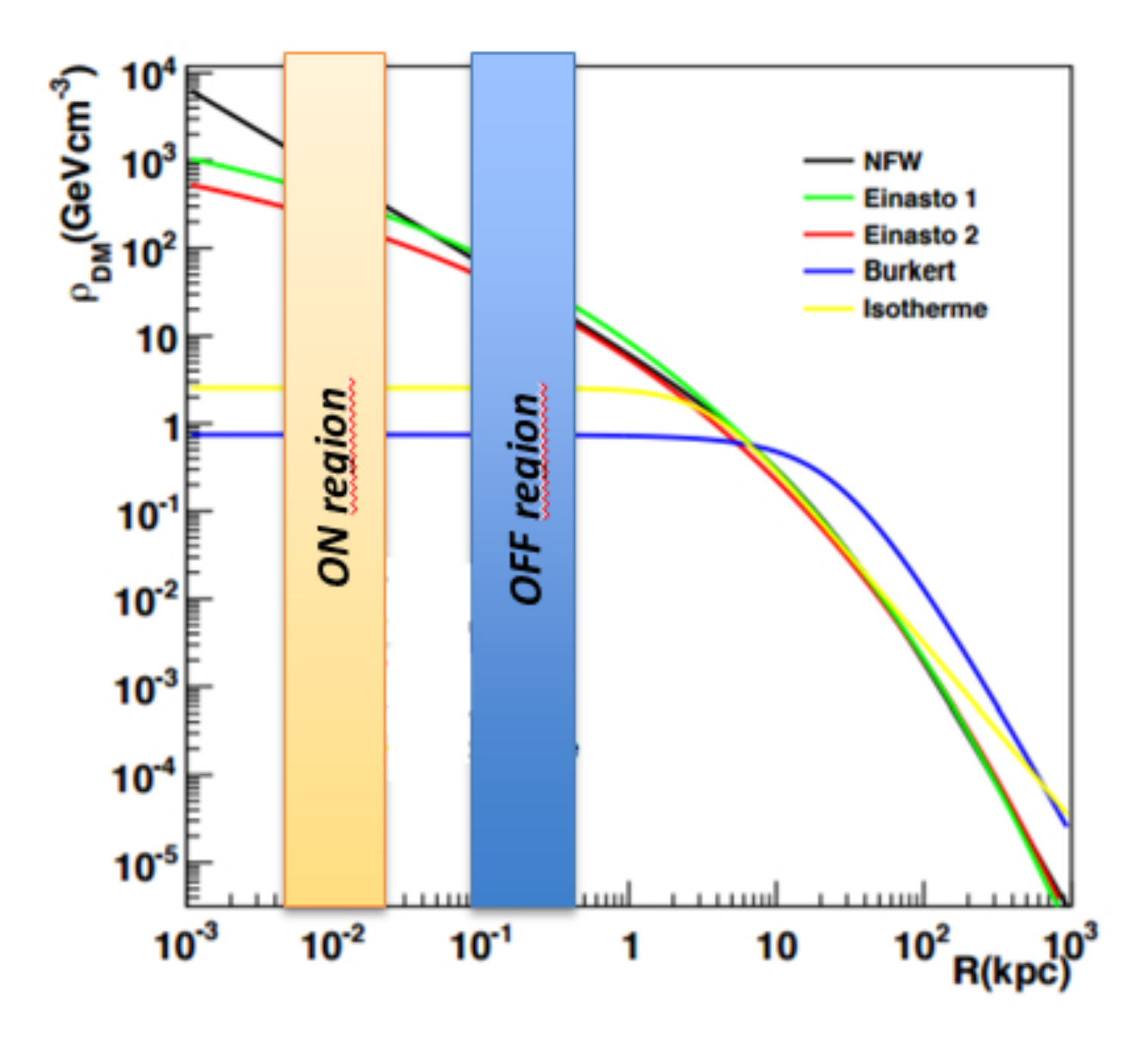}
        \caption{}
	\label{fig:DMprofile}
    \end{subfigure}%
    ~ 
    \begin{subfigure}[t]{0.45\textwidth}
        \centering
        \includegraphics[height=2.1in]{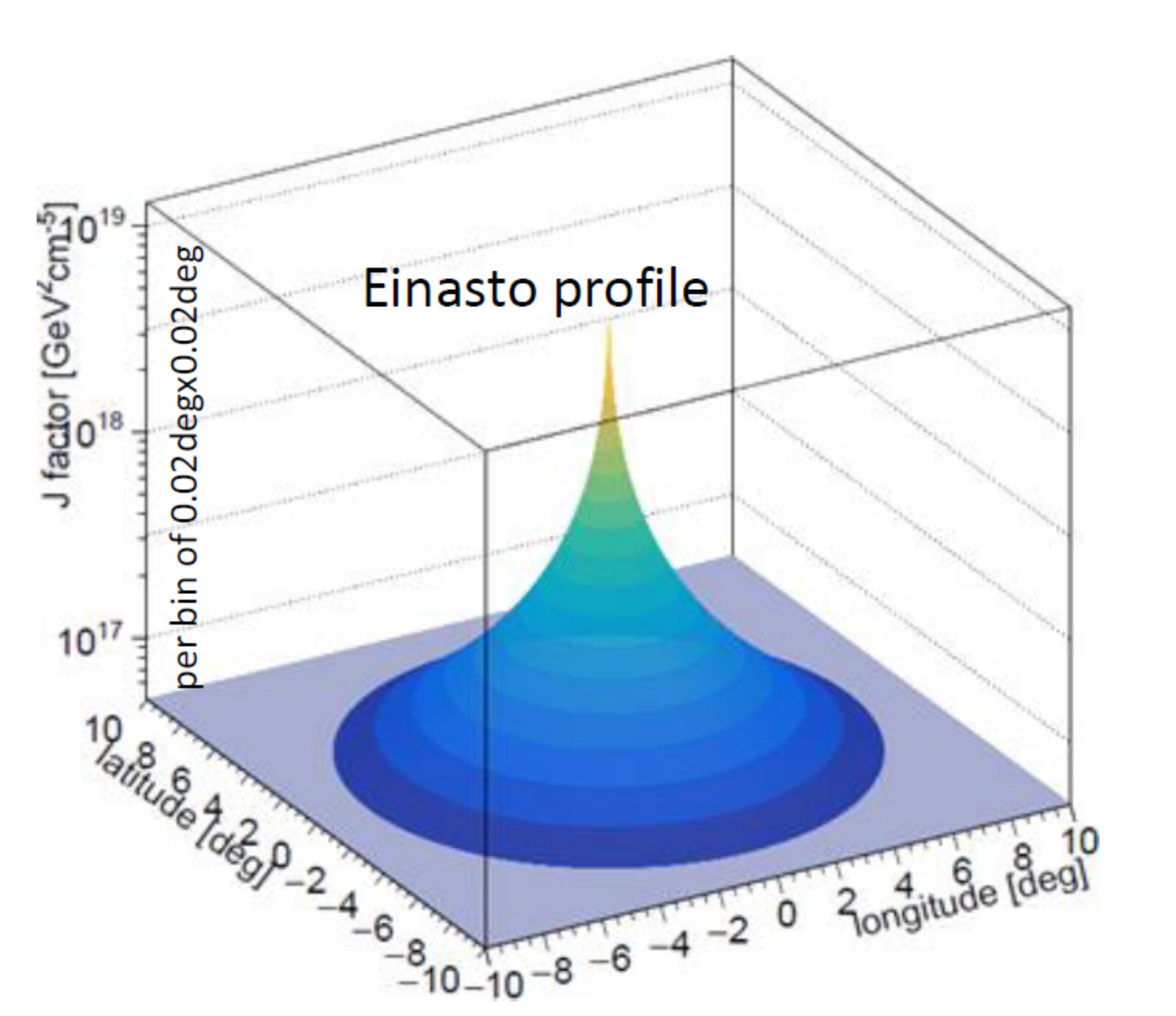}
        \caption{}
	\label{fig:JEinasto}
    \end{subfigure}
\caption{Dark matter density profiles in the Milky Way. {\it Left:} Examples of dark matter profiles in function of the distance from the GC. The two bands, respectively yellow and blue, represent the ON and OFF regions. {\it Right:} Mapping of the J-factor for the Einasto profile in galactic longitude and latitude per bin of $0.02^{\circ}\times0.02^{\circ}$. The profile is strongly peaked at the GC.}
\end{figure*}
cored profiles ({\it e.g.} Burkert profile), which are flat at distances below 1 kpc from the GC, and cuspy profiles ({\it e.g.} Einasto and NFW profiles) which are peaked toward the GC. One of these lasts is used for the following analysis. The Einasto profile is parametrized as 
$$\rho_{\rm Ein}(r)=\rho_s\exp{\bigg[\frac{-2}{\alpha}\bigg(\Big(\frac{r}{r_s}\Big)^{\alpha}-1\bigg)\bigg]}$$
with parameters extracted from in the previous publications \cite{bib:2011,bib:2013}.
Fig. \ref{fig:JEinasto} shows the J-factor values mapped in Galactic latitude and longitude for the Einasto profile. 
On the other hand, the residual background has a different spatial distribution with a nearly isotropic distribution. Thus, the specific spatial behavior of DM versus background can be exploited to improve the signal-to-background discrimination.

\subsection{Astrophysical targets for dark matter searches with VHE gamma-rays}\label{subsec:fig}

VHE $\gamma$-rays from DM annihilation can be searched for in different astrophysical targets: the Galactic Center which hosts the supermassive black hole Sagittarius A*, the central region of the Galactic Halo, dwarf galaxies satellites of the Milky Way, and nearby galaxy clusters. The most promising targets for DM searches with VHE $\gamma$-rays are the dwarf galaxies and the inner Galactic Halo. Dwarf galaxies are the most DM-dominated systems in the Universe and very low standard VHE astrophysical emission is expected. Tens of them have been recently discovered in the inner 100 kpc from the Galactic Center \cite{bib:dwarf}.  On the other hand, 
thanks to the proximity of the GC ($\sim8.5$ kpc), higher DM signal is expected from the central region of the Galactic Halo. However, searches there face significant standard VHE astrophysical emission, such as the Galactic diffuse background detected by the Fermi satellite \cite{bib:emissionFERMI}. Despite the presence of several standard astrophysical emissions, the Inner Galactic Halo region is the most promising place for DM detection. 

\section{Data analysis of the Galactic Centre observations with H.E.S.S. I}

\subsection{Observational dataset and definition of the region of interest}

The following analysis has been performed using the full 10-years data set of H.E.S.S. GC observations, collected during the first phase from 2003 to 2014 with the 4 telescopes of 12 m diameter and field of view (FoV) of $5^\circ$. A total of 254 hours of observations is available at the position of the GC, which corresponds to a doubled statistics with respect to the data set used in previous DM studies in 2011 \cite{bib:2011} and 2013 \cite{bib:2013}. The zenith angle of the observations never outmatches $50^\circ$ and
\begin{figure}[htbp] 
  \begin{center}
    \includegraphics[width=0.75\textwidth]{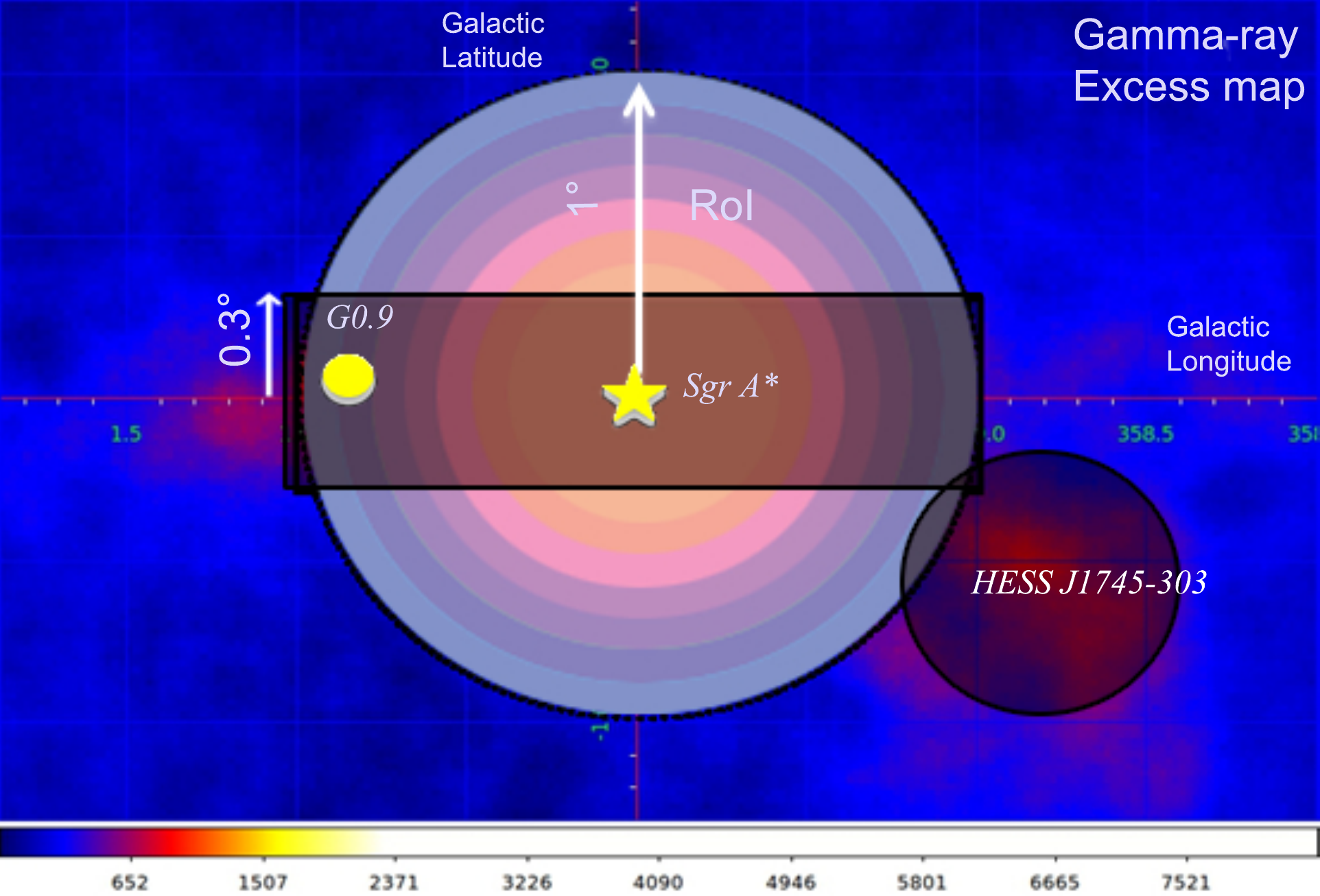}
    \caption{VHE $\gamma$-ray excess map in Galactic coordinates of the inner 300 pc of the Milky Way seen by H.E.S.S. Bright sources and diffuse emission are observed. The ON region for DM search is defined as a circle of $1^\circ$ centered at the GC and it is divided in 7 sub-regions of interest. The band of $\pm0.3^{\circ}$ around the Galactic Plane is excluded for the data analysis. A circle of radius $0.4^\circ$ centered on HESS J1745-303 is excluded as well.}
    \label{fig:GC}
  \end{center}
\end{figure} 
 it is on average of about $19^\circ$. The pointings are at a distance from GC between $0.5^\circ$ and $1.5^\circ$.\\
The DM signal is searched for in the ON region, referred to as the region of interest (RoI), which is defined as a circle of radius $1^\circ$ around the Galactic Center. This region harbors several VHE $\gamma$-rays sources such as HESS J1745-290 coincident with the supermassive black hole Sgr A$^*$ \cite{bib:source290}, G09+0.1 \cite{bib:sourceG09},  HESS J1745-303 \cite{bib:source303}, that contaminate the signal region with standard astrophysical emission. Diffuse emission was also detected by H.E.S.S. in the inner 300 pc of the Milky Way \cite{bib:diffuse} spatially correlated to the massive clouds of the central molecular zone. To avoid standard astrophysical contamination in both the signal and the background regions a band of $\pm$0.3$^\circ$  around the Galactic plane is excluded. A circle of radius $0.4^\circ$ centered on the position of the source HESS J1745-303 is also excluded. The ON region is divided in 7 subregions of width 0.1$^\circ$ which coincide with the spatial binning that is implemented in the likelihood analysis in Sec. \ref{sub:likelihood}. Fig. \ref{fig:GC} shows the $\gamma$-ray excess map of the inner 300 pc of the GC with overlapped the sub-RoIs used for the data analysis and the exclusion regions.

\subsection{VHE gamma-ray background measurement in the GC region}\label{subsec:final}

For each RoI corresponding to the ON region, the OFF region is defined for each observation, referred to as a {\it run}.
\begin{figure*}[htbp]
    \centering
    \begin{subfigure}[t]{0.45\textwidth}
        \centering
        \includegraphics[height=2.3in]{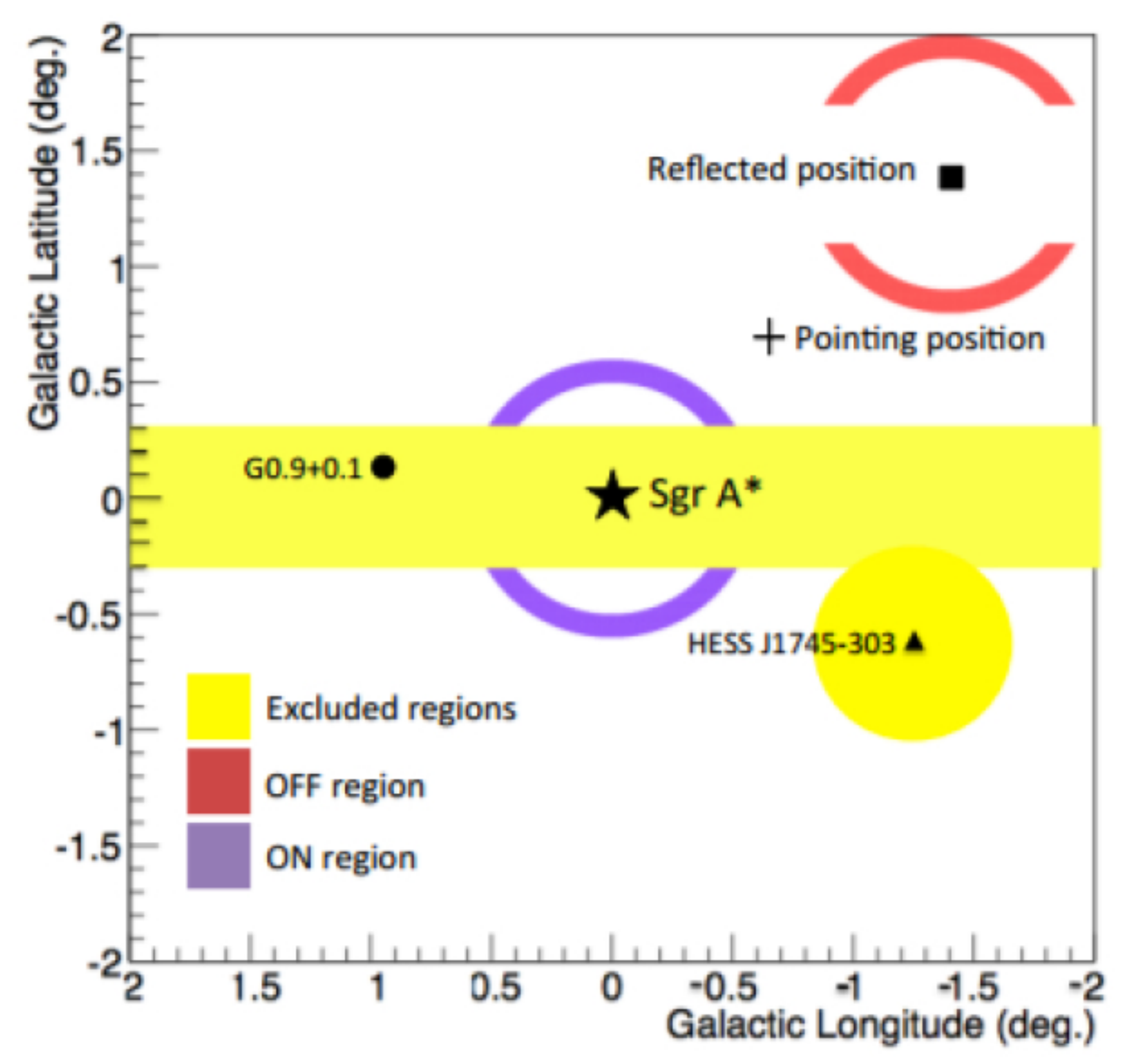}
        \caption{}
	\label{fig:OFF}
    \end{subfigure}%
    ~ 
    \begin{subfigure}[t]{0.45\textwidth}
        \centering
        \includegraphics[height=2.6in]{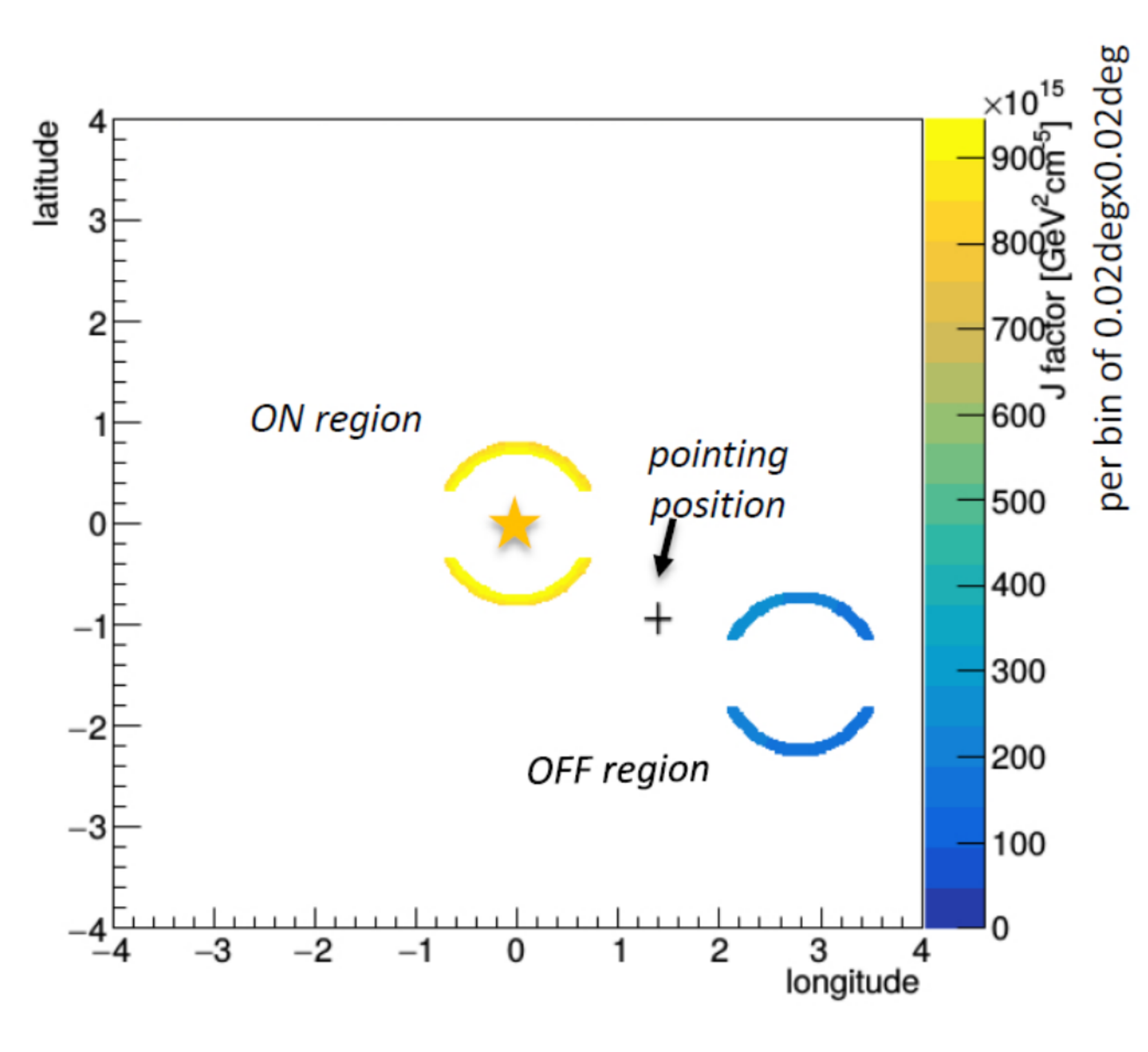}
        \caption{}
	\label{fig:JOFF}
    \end{subfigure}
\caption{Definition of the ON and OFF regions for a given pointing position (black cross) and one RoI. {\it Left:} The ON region is represented  for the sub-RoI (purple annulus) and the correponding OFF region (red annulus). The exclusion regions (yellow band and circle) are shown. {\it Right:} Projection of the map of the J-factors per bin size of $0.02^\circ\times0.02^\circ$ for the Einasto profile in the ON (yelllow) and OFF (blue) regions, respectively. }
\end{figure*} 
Indeed, for each run the OFF region is built symmetrically to the ON region with respect to the pointing position: this background extraction method is referred to as the reflected background method.
Fig. \ref{fig:OFF} shows an example for one RoI and one observation. The procedure is repeated for the 7 RoIs and the 609 observation runs. The obtained ON and OFF regions are thus taken with the same acceptance due to azimuthal symmetry, and have same shape and solid angle size.
The ON and the OFF regions host different amounts of DM, hence they have different J-factors. Their values are computed accordingly to the same reflected method.
Assuming a cuspy DM profile, the DM density gradient between the ON and OFF regions is strong enough to define them in the same observational field of view, {\it i.e.} with the same observation conditions.
Fig. \ref{fig:DMprofile} shows an example of the ON (yellow) and OFF (blue) annulii overlaid to the DM density profile. Fig. \ref{fig:JOFF} shows the projection of  the J-factor values of the Einasto profile on the ON and OFF regions for a specific observation and for one RoI. In this example, there is a DM gradient of a factor about 4 between the ON region (yellow annulus) and the OFF region (blue annulus). The DM signal is not null in the OFF region and it is accounted for in the likelihood data analysis (see Sec. \ref{sub:likelihood}).
In the definition of the OFF regions and the computation of the J-factors it is important to reject the exclusion regions and their reflected regions, both when they intersect the ON and the OFF regions. For some pointings very close to the GC, there are overlappings of the ON and OFF regions and areas where the J-factor value is larger in the OFF than in the ON region. These parts of the ON and OFF regions are also symmetrically discarded  in order to maintain the strong DM gradient required between them. The J-factor values computed for the single run are weighted by the live time of each observation to obtain the total J-factor for each RoI.

\subsection{The 2D-likelihood based analysis method}\label{sub:likelihood}

The data analysis method is based on a likelihood ratio test. The likelihood function is binned in 2 dimensions: energy (bins j) and space (RoIs, bins i) in order to exploit the different spectral and spatial behaviors of signal and background. The 2D-binned likelihood function writes as the product of the Poisson terms for the ON and OFF regions:
$$\mathcal{L}_{\rm ij}(\mathbf{N_{\rm ON}},\mathbf{N_{\rm OFF}}|\mathbf{N_{\rm S}},\mathbf{N_{\rm S}'},\mathbf{N_{\rm B}})=\frac{(N_{\rm S}+N_{\rm B})^{N_{\rm ON}}}{N_{\rm ON}!}e^{(N_{\rm S}+N_{\rm B})}\frac{(N_{\rm S}'+\alpha N_{\rm B})^{N_{\rm OFF}}}{N_{\rm OFF}!}e^{(N_{\rm S}'+\alpha N_{\rm B})}.$$
The factor $\alpha$ is defined as the ratio between the size of the OFF and ON regions. Here, $\alpha$ is equal to 1 by construction of the ON and OFF regions.
The values $N_{\rm ON}$ and $N_{\rm OFF}$ represent the number of photons measured in the ON and OFF regions, respectively -in each bin (i,j). The parameters $N_{\rm S}$ and $N_{\rm S}'$ represent the DM signal expected in the ON and OFF regions, respectively. The background expected in the ON region, $N_{\rm B}$, is computed from $d\mathcal{L}_{\rm ij}/dN_{\rm B}=0$.\\
The total likelihood is the sum over the individual likelihood $\log\mathcal{L}_{\rm ij}$ in the bin $(i,j)$ and it is used to build a test statistics. The likelihood ratio test $TS$ writes as:
$$TS=-2\log\Bigg(\frac{\mathcal{L}_{\rm W}}{\mathcal{L}_{\rm WO}}\Bigg),$$
where $\mathcal{L}_{\rm W}$ corresponds to the likelihood in presence of DM, and $\mathcal{L}_{\rm WO}$ the likelihood in the null hypothesis, {\it i.e.} without DM where the terms $N_{\rm S}$ and $N_{\rm S}'$ vanish.
In case no significant excess is observed in the ON region with respect to the OFF, upper limits at $95\%$ C.L. 
on $\langle \sigma v \rangle$ can derived for each DM mass by setting $TS=2.71$.

\section{Results}
No significant $\gamma$-ray excess is observed in any of the 7 RoIs.
Updated limits have been set on the annihilation cross section for the continuum and $\gamma$-line signals, using the new analysis method described above and the full data set of H.E.S.S.-I observations towards the GC.

\subsection{Continuum signal}
\begin{figure*}[htbp]
    \centering
    \begin{subfigure}[t]{0.45\textwidth}
        \centering
        \includegraphics[height=2.4in]{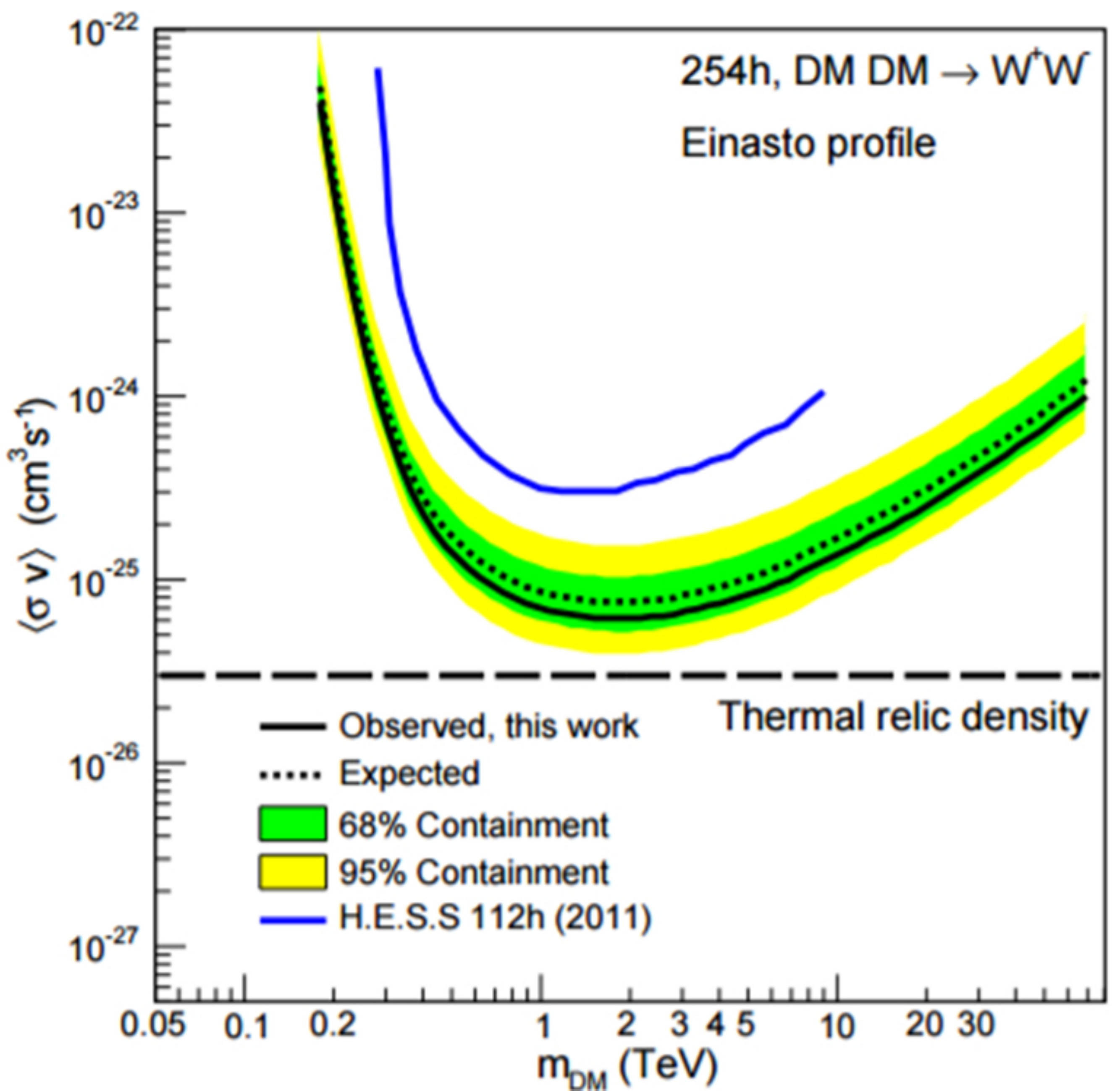} 
        \caption{}
	\label{fig:chW}
    \end{subfigure}%
    ~ 
    \begin{subfigure}[t]{0.45\textwidth}
        \centering
        \includegraphics[height=2.4in]{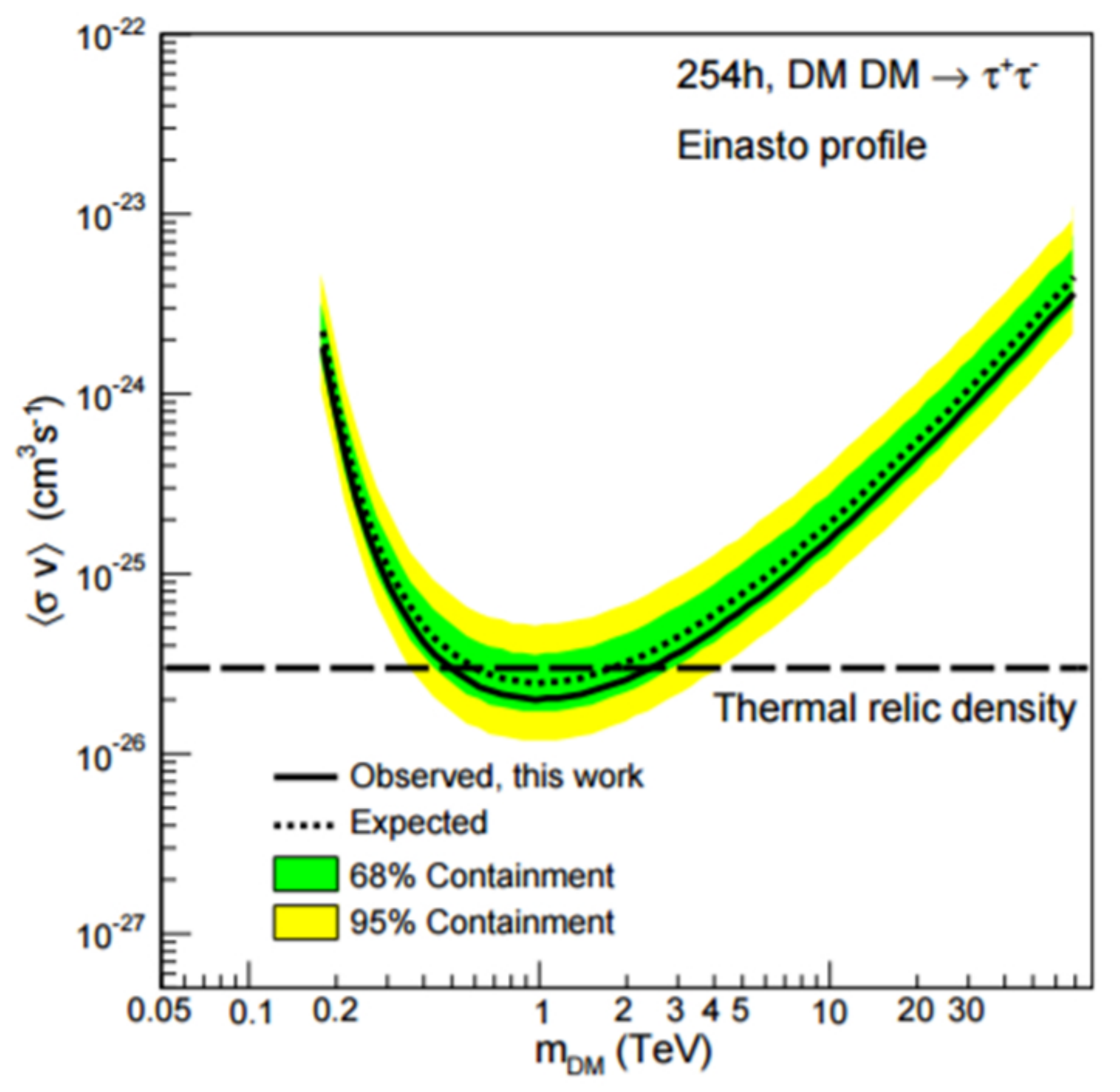}
        \caption{}
	\label{fig:chTau}
    \end{subfigure}
\caption{Constraints on the velocity-weighted annihilation cross section in the continuum channel: observed limits (solid black line) and mean expected limit (dotted black line) are shown together with the containment bands at 68\% (green band) and 95\% (yellow band). The horizontal black long-dashed line corresponds to the thermal relic cross section. {\it Left:} Limits in the $W^+W^-$ channel. The best limit is of $\sim6\times10^{-26}$ cm$^3$s$^{-1}$ at 1.5 TeV. The previous results from 2011 (112 h live time) are shown in blue. With the new limits an improvment of about 5 at 1 TeV is observed. {\it Right:} Limits in the $\tau^+\tau^-$ channel. The best limit is of $\sim2\times10^{-26}$ cm$^3$s$^{-1}$ at 800 GeV. For the first time, in this channel, H.E.S.S. was able to probe the thermal relic cross section.}
\label{fig:continuum}
\end{figure*}

The latest limits on the continuum are shown in Fig. \ref{fig:continuum}. In a black solid line in Fig. \ref{fig:chW} are the observed limits for the $W^+W^-$ channel.  
The threshold effect at low energies and the behavior like
$m_{\rm DM}^2$ at high masses are visible. These results are compatible with 
the mean expected limits (dotted black line), plotted with their $95\%$  (yellow band) and $68\%$ (green band) containment bands, respectively. 
The new best limits are of $\sim6\times10^{-26}$cm$^3$s$^{-1}$ at 1.5 TeV in $W^+W^-$ channel and of $\sim2\times10^{-26}$cm$^3$s$^{-1}$ at 800 GeV in $\tau^+\tau^-$ channel, respectively.
An improvement of a factor about 5 is obtained with respect to the limits previously published in 2011 \cite{bib:2011} (blue solid line), thanks to higher statistics, the 2D-binned likelihood analysis method and the introduction of the EW corrections to the $W^+W^-$ theoretical spectrum. The results on the annihilation in $\tau^+\tau^-$ are shown in Fig. \ref{fig:chTau} . Due to the harder spectrum of this channel compared to the $W^+W^-$ the best limits are shifted to a lower DM mass value. For the first time the thermal relic cross section (black dashed line) can be probed with H.E.S.S. observations.

\subsection{Line signal}

\begin{figure*}[htbp]
    \centering
    \begin{subfigure}[t]{0.45\textwidth}
        \centering
        \includegraphics[height=2.4in]{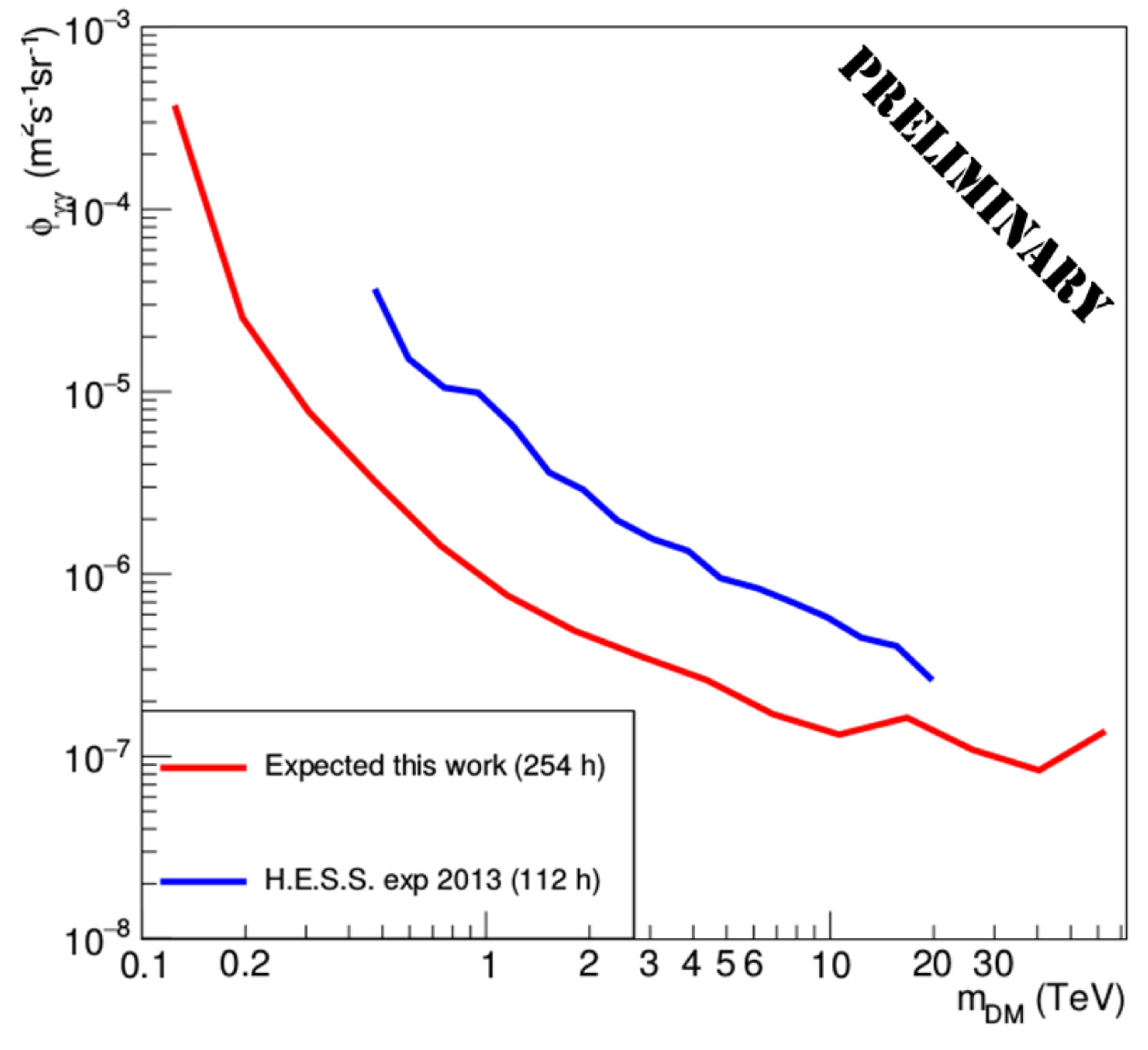}
        \caption{}
	\label{fig:limitFlux}
    \end{subfigure}%
    ~ 
    \begin{subfigure}[t]{0.45\textwidth}
        \centering
        \includegraphics[height=2.4in]{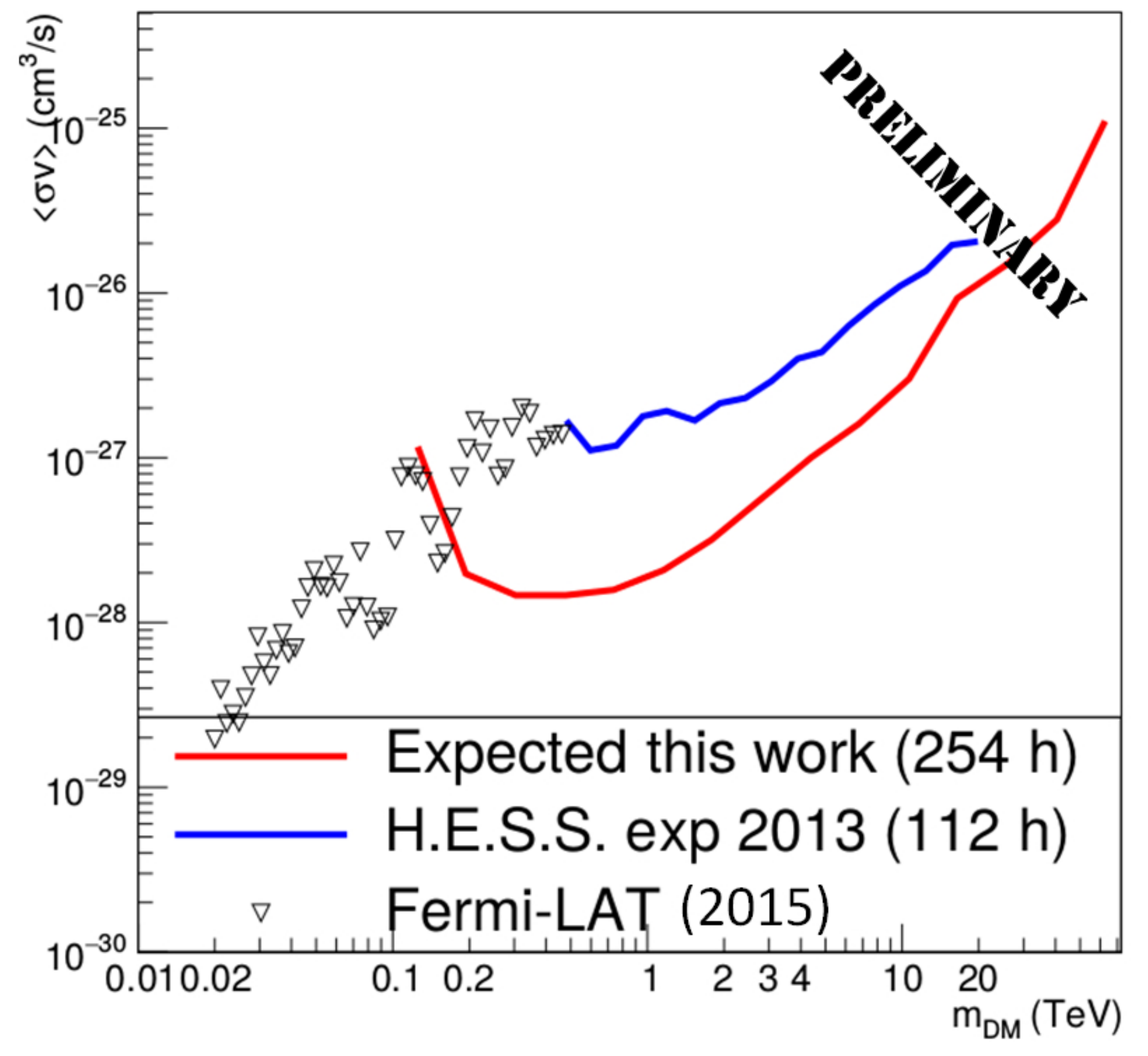}
        \caption{}
	\label{fig:limitSigma}
    \end{subfigure}
\caption{Mean expected limits of the prompt photons channel for this work (red solid line) and 2013 (blue solid line). {\it Left:} Limits on the annihilation photons flux. An improvement of a factor about 7 is observed. {\it Right:} Limits on the velocity-weighted annihilation cross section. Fermi limits from 2015 are also shown (black triangles).}
\label{fig:limits}
\end{figure*}

Updated limits have been computed in the prompt $\gamma\gamma$ channel. The same H.E.S.S.-I data set and 2D-binned likelihood approach have been used, as for the continuum analysis.
Here, a $\gamma$-ray like line has been produced starting from the annihilation spectrum described as a delta function $\frac{dN}{dE'}(E')=2\delta(E'-m_{\rm DM})$ centered at the DM mass.
The number of photons from DM annihilation in each RoI and energy bin is computed by
$$N_{\gamma}(E)=\frac{1}{4\pi}\frac{\langle\sigma v\rangle}{2m_{\rm DM}^2}\int\frac{dN}{dE'}(E')R(E,E')T_{\rm obs}A_{\rm eff}(E')dE'\times J(\Delta\Omega).$$
The input DM spectrum is spread with a Gaussian energy resolution $R(E,E')=\frac{1}{\sqrt{2\pi}\sigma}e^{-\frac{(E-E')^2}{2\sigma^2}}$ of variance $\sigma/E=10\%$. This value is almost independent on zenith angle and offset of the observation. $A_{\rm eff}$ is the energy-dependent effective area and  $T_{\rm obs}$ the observation time.\\
The expected limits are computed by setting the condition $N_{\rm ON}\equiv N_{\rm OFF}$, {\it i.e.} assuming no excess in the ON region. Fig. \ref{fig:limitFlux} shows the updated expected limits on the $\gamma$ flux from DM annihilation (red solid line) compared to the previous ones (blue solid line) published in 2013 \cite{bib:2013}. 
The limits on the DM prompt annihilation cross section in Fig. \ref{fig:limitSigma} show the threshold effect at low mass and the behavior like $m_{\rm DM}^2$ at high mass, as for the continuum.
The doubled statistics with the new observations is expected to improve the limits of a factor $\sim1.4$. In addition, as for the continuum, the new improved 2D analysis is applied and the background Poisson term has been implemented in the likelihood. Moreover, a new raw data analysis is used allowing for improved sensitivity at energies lower than $\sim1$ TeV. Indeed, the energy threshold is lowered and the best limits are shifted to smaller values. The DM mass range is also increased from 200 GeV to 70 TeV.
At 1 TeV a factor of improvement of about 7 is obtained. A factor $\sim4$ is explained by the optimized statistical method and the increased observation time. The remaining improvement comes from the use of the new raw data analysis which is more sensitive than the old one below $\sim1$ TeV, while they are compatible at high energies.
Above about 300 GeV these new limits surpass those from Fermi-LAT \cite{bib:Fermi} of a factor $\sim4$. 

\section{Summary}

The GC region has been observed with the first phase  of H.E.S.S. to search for a signal of self-annihilation of DM particles with 10 years of operation. The data set amounts to 254 hours of observation at the nominal position of Sgr A$^*$. 
No significant excess is found in the regions of interest and new stronger constraints are set on the DM annihilation cross section, the strongest so far in the TeV DM mass range.
The distinct spectral and spatial features of the DM signal have been exploited for a further discrimination against the residual background. Higher $\gamma$-ray statistics, improved new raw data analysis and optimized 2D-likelihood method enable significant improvement over the previous results.
The updated expected $95\%$ C.L. upper limits improve the previous ones of a factor $\sim5$ in the continuum analysis for the $W^+W^-$ channel. In the $W^+W^-$ channel a best limit of $\sim6\times10^{-26}$cm$^3$s$^{-1}$ is reached at $\sim1.5$ TeV. For the first time H.E.S.S. was able to probe the thermal relic cross section in the $\tau^+\tau^-$ channel with a best limit of $\sim2\times10^{-26}$cm$^3$s$^{-1}$ at $\sim800$ GeV.
Preliminary results on the DM $\gamma$-line signal show an improvement of a factor $\sim7$ at $\sim1$ TeV.
The expected limits significantly surpass the Fermi-LAT limits around 300 GeV.
Larger statistics and optimized observational strategy of the GC with H.E.S.S. II pointings  up to $3^\circ$ in Galactic latitudes will improve further the sensitivity to DM annihilations. These data analyses pave the way to future studies with CTA observations, that will provide higher sensitivity compared to H.E.S.S. \cite{bib:CTA0,bib:CTA}. 

\end{document}